\documentclass[11pt]{article}
\usepackage{moriond,epsfig}

\def\be{\begin{equation}}
\def\ee{\end{equation}}
\def\bea{\begin{eqnarray}}
\def\eea{\end{eqnarray}}

\begin{document}
\begin{flushright}ULB-TH/09-13\end{flushright}
\vspace*{4cm}
\title{Can LHC disprove Leptogenesis ?}

\author{ G.~Vertongen }

\address{Univertsit\'e Libre de Bruxelles, Service de Physique Th\'eorique CP 225,\\ Blvd du Triomphe, 1050 Bruxelles, Belgium.}

\maketitle
\abstracts{
Among the mechanisms which successfully explain the generation of the Baryon Asymmetry of the Universe, Leptogenesis through right-handed neutrino decays is especially attractive. Unfortunately, this theory suffers from a lack of testability. Indeed, the high energy relevant ingredients in the asymmetry creation are either indirectly linked to low energy observables or unreachable by our present experiments. We propose here to take the problem the other way around by studying whether this mechanism could at least be disproved. We argue that the observation of a right handed gauge boson $W_R$ at future colliders could play this role. 
}

\section{Introduction}
Considering right-handed neutrinos $N$ in addition to - or being part of - the Standard Model offers the possibility to generate neutrino masses through the well-known see-saw mechanism. The latter finds its origin in the Yukawa interactions and the Majorana masses of the right-handed neutrinos
\begin{equation}
{\cal L} \owns
- \overline{L} \,{\widetilde H} \, {Y_\nu^\dagger}  \, N
-\frac{1}{2}\,\overline{N} \, {m_N} \,{{N}^c}
+\mbox{h.c.}
\end{equation}
where $L$ stands for the lepton weak doublets and $\tilde H$ is related to the standard Brout-Englert-Higgs (hereafter simply Higgs) doublet $H \equiv (H^+, H^0)$ by $\tilde H = i \tau_{2} H^*$.  This is not the only virtue of the see-saw mechanism : it also provides a way to generate the baryon asymmetry of the Universe (BAU) through the so-called Leptogenesis mechanism \cite{fy}, and this without requiring any further interactions. 

In the standard scenario, the resulting baryon asymmetry could be parametrised by the product of three quantities : $\epsilon_N$, the amount of CP asymmetry created through right-handed neutrino decay, the Boltzmann equations which determine  the efficiency $\eta$, and finally $r_{{\cal L}\rightarrow {\cal B}}$, the rate of $L$ to $B$ sphaleron conversion (-28/79 in the SM \cite{Luty:1992un}) :
\begin{eqnarray}
Y_{\cal B} = \epsilon_N\,\eta\,Y_N^{eq}(T\gg m_N)\,r_{{\cal L}\rightarrow {\cal B}}
\end{eqnarray}
with $Y_i\equiv n_i/s$, $Y_{\cal B}\equiv Y_{B}-Y_{\bar{B}}$, $Y_{\cal L}\equiv Y_L-Y_{\bar{L}}$, $n_i$ the comoving number density of the species "i", "eq" referring to the equilibrium number density,  and $s$ the comoving entropy density.

\bigskip
Although very elegant, Leptogenesis present a major practical deficiency since it resists to any tentative of being tested.
For example, when considering a hierarchical spectrum for the right-handed neutrinos, oscillations parameters constraint the right-handed neutrinos to have a mass far above our current accelerator limits \cite{di}. This scale could however be lowered by invoking resonant Leptogenesis mechanism \cite{lowscale}, \textit{i.e.} by considering a nearly degenerate spectrum ; unfortunately the production of $N$ generically suffers in this case from a suppression of the Yukawa couplings coming from the neutrino mass constraints.
On the other hand, when one tries to relate the CP violation which generate the asymmetry with low energy observables, one can show that there is no direct link between them since Leptogenesis depends generically on high energy parameters which do not enter in the measurable low energy observables (in particular in the neutrino masses).

\bigskip
Here, we propose to tackle the problem the other way around : if Leptogenesis could not be \textit{tested}, could it be at least \textit{disproved} ? We propose here one possibility, namely through the observation of a right-handed gauge boson $W_R$. 
Unlike in the standard case where right-handed neutrinos are introduced in an isolated way, these new fields naturally take place in higher representations of Grand Unified Theories, where they live together with SU(2)$_R$ gauge bosons. It must be stressed that neglecting, as usually done, the effect of SU(2)$_R$ gauge bosons is an assumption, as it has been shown that their effect must be taken into account if the $W_R$ mass is not several orders of magnitude above the $N$ ones \cite{Carlier}.


\section{Leptogenesis in the presence of a $W_R$}
The relevant interactions when introducing right-handed gauge bosons are
\begin{equation}
{\cal L} \owns \frac{g}{\sqrt{2}}
W_R^{\mu} \left( \bar{u}_R \gamma_{\mu} d_R + \bar{N} \gamma_{\mu} \,l_R \right)
\end{equation}
where $N$ and the right-handed charged leptons ($l_R=e_R,\mu_R,\tau_R$), and $u_R$ and $d_R$,  are members of a same $SU(2)_R$ doublet. They induce two types of interactions which dramatically influence the asymmetry generation.

\subsection{Decay effects}
The first effect induced comes from the presence of new decay channels. Depending on the relative values of the $W_R$ and the right-handed neutrino masses, the latter could decay in two or three bodies. Given the large value of the gauge couplings, the three body decay can easily compete with the Yukawa two body decay. Since these new channels are CP conserving, they introduce a dilution of the asymmetry $\epsilon^{(0)}$ generated in the standard case, \textit{i.e.} without $W_R$ effects:
\begin{small}
\begin{eqnarray}
\epsilon = \frac{\Gamma_N ^{(l)} - \overline{\Gamma}_N ^{(l)}}{\Gamma_{tot}^{(l)} + \Gamma_{tot}^{(W_R)}} \equiv \epsilon^{(0)} \frac{\Gamma_{tot}^{(l)}}{\Gamma_{tot}^{(l)} + \Gamma_{tot}^{(W_R)}} \,,
\end{eqnarray} 
\end{small}
where $\Gamma^{(l)}$ and $\Gamma^{(W_R)}$ stand for decays through Yukawa and gauge couplings respectively. Unlike in standard Leptogenesis where the efficiency of the asymmetry creation could be as high as order 1, the dilution effect induced by right-handed gauge bosons leads automatically to the following upper bound
\begin{small}
\begin{eqnarray}
\eta \leq \eta^{\mbox{max}} = \frac{\Gamma_{tot}^{(l)}}{\Gamma_{tot}^{(l)} + \Gamma_{tot}^{(W_R)}} \,.
\end{eqnarray}
\end{small}

\subsection{Scattering effects}
New scatterings are also entering the game and, depending whether or not a $W_R$ is present as an external particle, could be distinguished in two categories. In this work, we only considered the leading effects, \textit{i.e.}  scatterings involving the $W_R$ as an internal particle. Indeed, since $W_R$ are strongly coupled to the thermal bath, the $W_R$ population remains in equilibrium down to very low temperatures, thus Boltzmann suppressing scatterings where $W_R$ is an external particles.

An interesting feature of the leading scatterings is their peculiar decoupling behaviour with temperature in the $Y_N$ Boltzmann equation. To evaluate the strength of an interaction on the evolution of a particle population, we must compare the corresponding interaction rates $\gamma_i$ not only with the Hubble expansion rate, but also with the particle number density concerned : $\gamma_i$ \textit{vs} $(n^{eq}_{{N}}\,H)$. Unlike in annihilation cases where the reaction rate $\gamma_i$ is twice Boltzmann suppressed, when only one external $N$ is present in a given state, $\gamma_i$ is Boltzmann suppressed only once at low temperatures, which compensate the Boltzmann suppression coming from $n_N^{eq}$. The decouplings of such scatterings are then only \textit{linearly} dependent on the temperature, making them very efficient down to low temperatures. Note that this linear decoupling of the $W_R$ interactions is also the origin of much stronger suppressions than from the $W_L$ interactions in other Leptogenesis frameworks, such as from the decay of a scalar fermion triplet \cite{typeIIleptoeffic}.

\subsection{Boltzmann equations}

\begin{figure}
\center
\includegraphics[width=8cm]{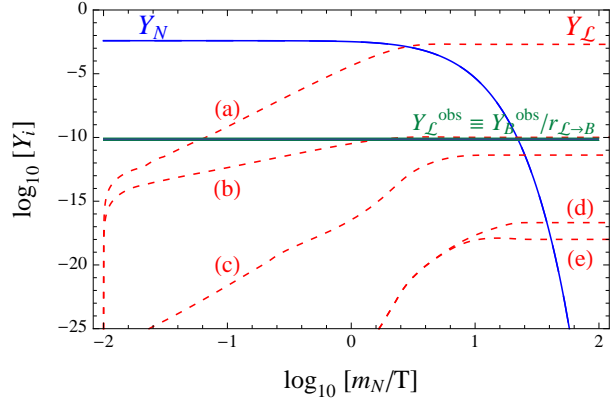}
\caption{Evolution of the abundances as a function of $z\equiv m_N/T$. Plain (Dashed) lines represents $Y_N$ ($Y_{\cal L}$) evolutions in the cases explained in the text. Green (grey) horizontal band represents the observed baryon asymmetry assuming sphaleron conversion.
\label{fig:evolution_abundances}}
\end{figure}

Adding new gauge interactions rates to the standard one, we get the following Boltzmann equations for the evolution of the comoving abundances as a function of $z\equiv m_N/T$:
\begin{small}
\begin{eqnarray}
zH(z)s\, Y'_{N} &=& -\left(\frac{Y_{N}}{Y_{N}^{\rm eq}}-1 \right) \left(\gamma_{N}^{(l)} + \gamma_{N}^{(W_R)} + 2 \gamma_{Hs} + 4\gamma_{Ht}+ 2 \gamma_{Nu} + 2 \gamma_{Nd} + 2 \gamma_{Ne} \right)\nonumber\\
& &- \left(\left(\frac{Y_N}{Y_N^{eq}}\right)^2 - 1 \right) \gamma_{NN}\label{NBoGauge}\\
zH(z)s\, Y'_{\cal L} &=&\gamma_{N}^{(l)} \epsilon_{N} \left(\frac{Y_{N}}{Y_{N}^{\rm eq}}-1\right) - \left(\gamma_{N}^{(l)}+ \gamma_{N}^{(W_R)}\right)\frac{Y_{\cal L}}{2\,Y_{L}^{\rm eq}}\nonumber\\
&&
-\frac{Y_{\cal L}}{Y_{L}^{\rm eq}}\left(2\,\gamma_{Ns}^{\rm sub}+2\,\gamma_{Nt}+2\,\gamma_{Ht} + 2\,\gamma_{Hs}\,\frac{Y_{N}}{Y_{N}^{\rm eq}}
+ \,\gamma_{Nu}
+ \,\gamma_{Nd}
+ \,\gamma_{Ne}\,\frac{Y_{N}}{Y_{N}^{\rm eq}}\right) \,\,\label{LBoGauge}
\end{eqnarray}
\end{small}
where the $'$ denotes the derivative with respect to $z$, and where $\gamma_N^{(W_R)}$ and $\gamma_{N,u,d,e}$ are the new decay channel and the new scatterings respectively. The other interactions rates are the standard one \cite{}. The resolution of these equations allows us to study the effects of gauge interactions on standard leptogenesis. To illustrate this, let us see how each new interaction affects the asymmetry creation by taking an example where $m_{W_R}=3$\,TeV, $m_N=500$\,GeV and $\tilde{m}=10^{-3}$\,eV. As well known, standard Leptogenesis successfully produces the BAU (case $a$ in fig.\ref{fig:evolution_abundances} ). Adding successively to this case the effect of the three body decay (case $b$)  and gauge scatterings (case $c$) in the $Y_N$ Boltzmann equation lead in the first case to a large dilution due to the relative strenght of gauge decays compared to the Yukawa decays, which is even more suppressed in the second case thanks to the strong thermalisation effect of gauge scatterings at high temperatures. These effects alone are sufficient to disprove Leptogenesis in this case. This effects are amplified when integrating the $W_R$ effects in the $Y_{\cal L}$ Boltzmann equation : gauge scatterings are putting more closer lepton from chemical equilibrium at high temperatures (case $d$), while three-body decays does only affects marginally the asymmetry production in this case  (case $e$). Altogether, we end up with an efficiency of $\eta\sim1.6\cdot10^{-18}$ which is far below the required value.

\section{Efficiencies results}
\begin{figure}
\center
\includegraphics[width=6.5cm]{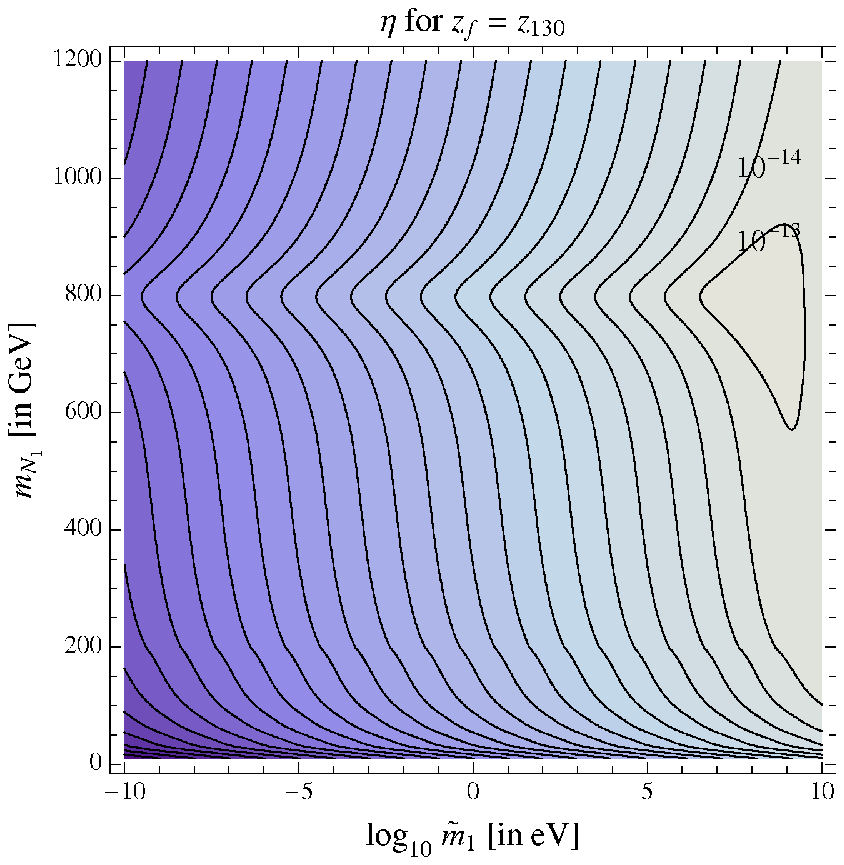}
\includegraphics[width=6.5cm]{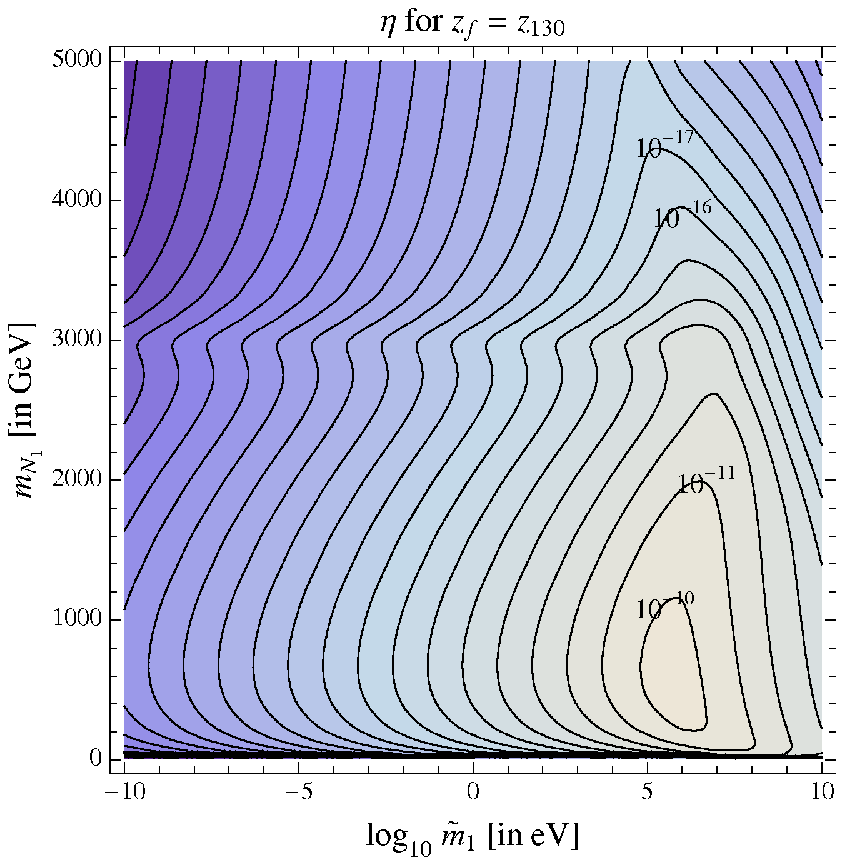}
\caption{Iso-efficiencies for (\textit{left}) $m_{W_R} = 800$\,GeV and (\textit{right}) $m_{W_R} = 3$\,TeV as a function of $\tilde{m}$ and $m_N$.
\label{fig:iso_efficiencies}}
\end{figure}

The resolution of the Boltzmann equations (\ref{NBoGauge})-(\ref{LBoGauge}) allow us to infer the efficiencies as a function of $m_{W_R}, m_N$ and $\tilde{m}$. In fig.\,(\ref{fig:iso_efficiencies}) are plotted the iso-efficiencies for $m_{W_R} = 800$\,GeV, the actual experimental lower limit \cite{PDG}, and for $m_{W_R} = 3$\,TeV, which is the mass that LHC will reasonably be able to probe \cite{LHCstudies}. In all cases, the resulting efficiency is far below $7\cdot10^{-8}$, which is the minimum value necessary to get the observed baryon asymmetry $Y_{\cal B} \simeq 8\cdot10^{-11}$.

In order to comment the evolution of the efficiency with $m_N$ and $\tilde{m}$, let us start from the highest efficiency value in the $m_{W_R} = 3$\,TeV case. If one decrease the value of $m_N$, the decoupling temperature also decreases linearly, leading to less time for the creation of an asymmetry since we assumed the sphaleron conversion to stop at $T\sim130$\,GeV (for $m_h \sim120$\,GeV) \cite{Burnier}. On the other hand, increasing the N mass will lead to an earlier decoupling, but this effect is somewhat preempted by the three body decay which becomes more efficient in this case. For $m_N \leq m_{W_R}$, the $Ne$ $s$-channel scattering gets suppressed, leading to a slight efficiency enhancement. Finally, the $m_N > m_{W_R}$ regime is always forbidden since in this case, the allowed two body decay lead to a strong dilution of the created asymmetry. Varying now $\tilde{m}$ down to low values, we suppress the strength of the washout scatterings involving neutrino yukawa couplings but also the strength of the CP violation source. In total, the efficiency gets worst. Increasing $\tilde{m}$ gives more CP violation, but renders the $\Delta L=2$ washout interaction, which evolves like $\tilde{m}^2$, more stronger, thus leading to a suppression of the efficiency. All in all, it turns out that the maximal efficiency we get is $\eta \simeq 10^{-10}$, which is reached for $\{m_N, \tilde{m}\} = \{600\,\mbox{GeV}, 10^6\,\mbox{eV}\}$.

\section{Bounds on m$_{W_R}$}
\begin{figure}
\center
\includegraphics[width=6.5cm]{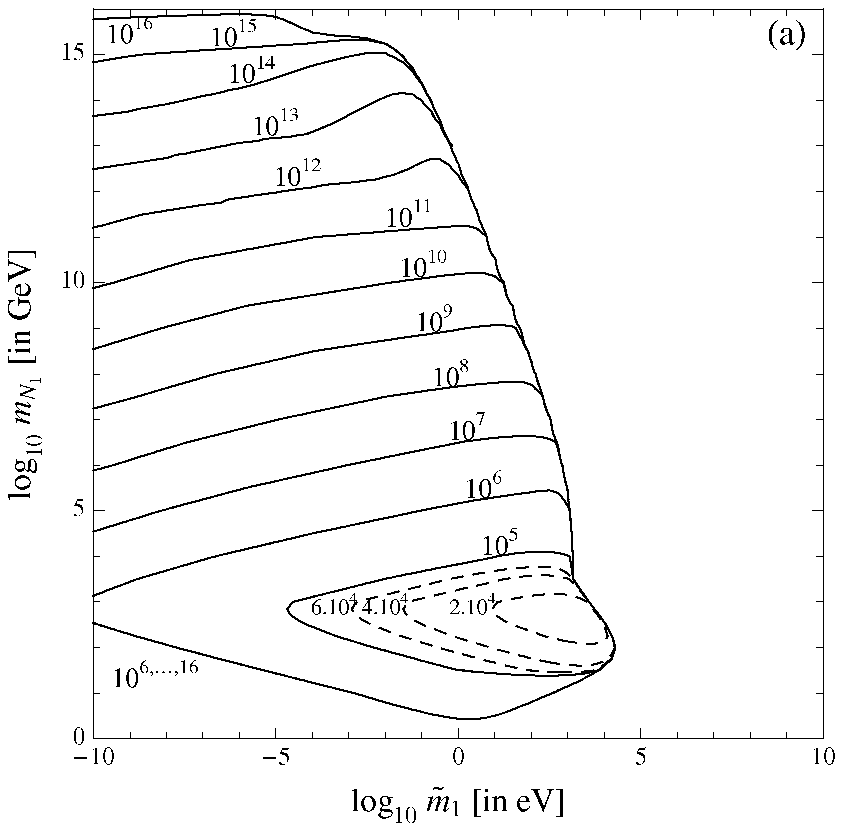}
\includegraphics[width=6.5cm]{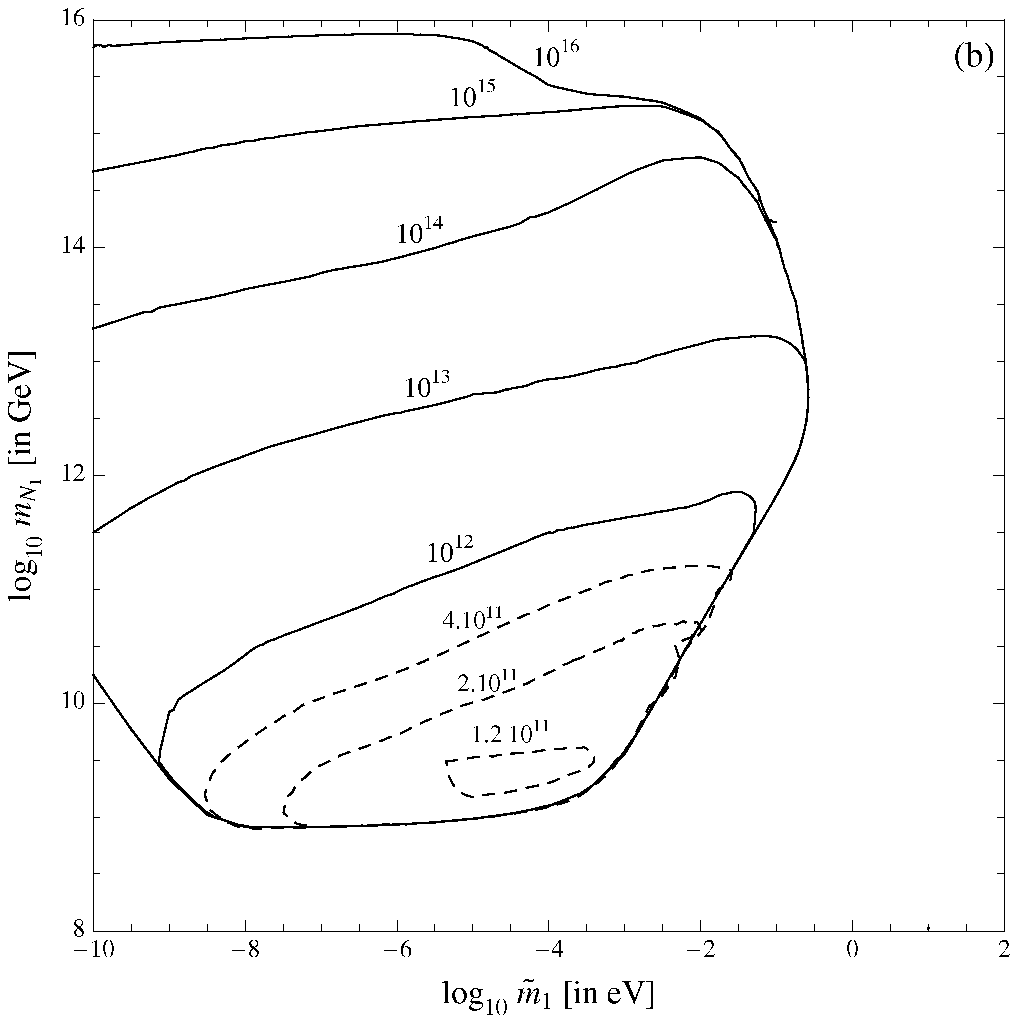}
\caption{For different values of the $W_R$ mass, the inner part of each curve represent the region for successful leptogenesis. \textit{Left (Right)} : $\epsilon = 1 \;( \frac{3}{16\, \pi}\, \frac{m_N}{v^2}\,\sqrt{\Delta m^2_{atm}})$. 
\label{fig:wr_limits}}
\end{figure}

As can be easily understood, increasing the value of the $W_R$ lead to less suppression, and thus allows at some point for a successful Leptogenesis. In fig.\ref{fig:wr_limits}, for each fixed value of $m_{W_R}$, the interior part of the contour represents the allowed $\{m_N, \tilde{m} \}$ region for a successful Leptogenesis. In the left panel, we set the CP violation parameter $\epsilon_N$ to 1, which, as previously stressed, represents the best configuration for a baryon asymmetry creation. In this case, the minimal value allowed for a successfull leptogenesis is $m_{W_R} = 18$\,TeV. Note that leptogenesis could not be achieved if $m_N$ is lighter than $2.6$\,GeV. The right panel represents the case where CP violation is resulting from the decay of hierarchical right-handed neutrinos \cite{di}:
\begin{eqnarray}
\epsilon_N < \frac{3}{16\, \pi}\, \frac{m_N}{v^2}\,\sqrt{\Delta m^2_{atm}}\,.
\end{eqnarray}
In this case, a lower bound of $m_{W_R} \simeq 10^{11}$\,GeV is obtained.

\section{Generalisation to several right-handed neutrinos}
In what precedes, we study the influence of a low scale $W_R$ on Leptogenesis induced by a \textit{single} right-handed neutrino. However, as explained in the introduction, a sufficient amount of CP asymmetry could only be produced through the well known resonant mechanism, \textit{i.e.} when at least two of the right-handed neutrinos present a nearly degenerate spectrum ; this means that the resolution of Boltzmann equations for at least 2 right-handed neutrinos would be mandatory. However, as shown in the Appendix A of ref. \cite{Frere:2008ct}, the asymmetry created by two right-handed neutrinos is bounded by the sum of both asymmetries we get in the single $N$ case, with $\tilde{m} = \tilde{m}_1$ and $\tilde{m} = \tilde{m}_2$ ($\tilde{m}_i$ referring to the value of $\tilde{m}$ of $N_i$). Our previous conclusion remains then unchanged.

In addition, when considering more right-handed neutrinos, a non-trivial flavour structure could take place in the Yukawa couplings. This would imply to study the generation of the asymmetry in each flavour separately, rather than the total lepton number as we did. From the disproving point of view, the worst situation is achieved when the $N$ creates an asymmetry in a flavour orthogonal to the one of its SU(2)$_R$ partner ; indeed, in this case gauge scatterings are only present in the $Y_N$ Boltzmann equation. However, as shown in the above example and more generally in ref. \cite{Frere:2008ct}, dilution through the $W_R$ is so strong that it is sufficient to rule out Leptogenesis also in this case.

\section{Conclusion}
We have shown that the discovery of a $W_R$ at LHC or future colliders would rule out the possibility of creating a sufficient lepton asymmetry from the decay of a right-handed neutrino. The presence of a $W_R$ induce new decay channels leading to strong dilution and washout of the created asymmetry, as well as strong scattering processes. Moreover, we showed that successful leptogenesis could be achieved in the resonant (hierarchical) case if the $W_R$ is heavier than 18 TeV ($10^{11}$\,GeV). Generalisation to other Leptogenesis models (type II \& III see-saw mechanisms) can also be found in ref. \cite{Frere:2008ct}.

\section*{Acknowledgments}
This work has been done in collaboration with J.~M.~Fr\`ere and T.~Hambye. The author received partial support from the Belgian Science Policy (IAP VI-11) as well as from the IISN.

\end{document}